\documentclass{rmaa}

\title{A comparison of the radio and optical time-evolution of HH~1 and 2}
\author
{L. F. Rodr\'\i guez
\affil{Instituto de Radioastronom\'\i a y Astrof\'\i sica, UNAM}
A. C. Raga
\affil{Instituto de Ciencias Nucleares, UNAM}
A. Rodr\'\i guez-Kamenetzky
\affil{Instituto de Radioastronom\'\i a y Astrof\'\i sica, UNAM and Instituto de Astronom\'\i a Te\'orica y Experimental, (IATE-UNC)}
C. Carrasco-Gonz\'alez
\affil{Instituto de Radioastronom\'\i a y Astrof\'\i sica, UNAM}
}

\fulladdresses{
\item C. Carrasco-Gonz\'alez: Instituto de Radioastronom\'\i a y Astrof\'\i sica, Universidad Nacional 
Aut\'onoma de M\'exico, A. P. 3-72 (Xangari), 58089 Morelia, Michoac\'an, M\'exico (c.carrasco@crya.unam.mx)
\item A. C. Raga: Instituto de Ciencias Nucleares,
Universidad Nacional Aut\'onoma de M\'exico, Ap. 70-543, 04510 Cd. Mx., M\'exico
(raga@nucleares.unam.mx)
\item L. F. Rodr\'\i guez: Instituto de Radioastronom\'\i a y Astrof\'\i sica, Universidad Nacional 
Aut\'onoma de M\'exico, A. P. 3-72 (Xangari), 58089 Morelia, Michoac\'an, M\'exico (l.rodriguez@crya.unam.mx)
\item A. Rodr\'\i guez-Kamentzky: Instituto de Astronom\'\i a Te\'orica y Experimental,
(IATE-UNC), X5000BGR C\'ordoba, Argentina (adrianark@gmail.com)
}

\shortauthor{Rodr\'\i guez et al.}
\shorttitle{Radio and optical time-evolution of HH~1 and 2}

\SetVolume{00} \SetFirstPage{000} \SetYear{2017}

\ReceivedDate{\today} \AcceptedDate{Year Month Day} 

\resumen{Presentamos una comparaci\'on entre la evoluci\'on temporal en los pasados
  $\sim 20$ a\~nos de la emisi\'on de radio continuo y de H$\alpha$ de HH~1 y 2. Encontramos
  que el radio continuo y H$\alpha$ de los dos objetos muestran evoluciones similares, con HH~1
  debilit\'andose y HH~2 volvi\'endose considerablemente m\'as brillante (aproximadamente un factor de 2). Tambi\'en encontramos
  que el cociente $F_{\rm H\alpha}/F_{ff}$ (entre H$\alpha$ y el radio continuo) de HH~1 y 2 tiene
  valores mayores que los encontrados t\'\i picamente en nebulosas planetarias (PNe), lo cual interpretamos como
  una indicaci\'on que la emisi\'on libre-libre y de H$\alpha$ de HH~1/2 es producida
  en regiones emisoras con temperaturas menores ($\sim 2000$~K) que la emisi\'on de las
  PNe (con $\sim 10^4$~K).}

\abstract{We present a comparison between the time-evolution over the past $\sim 20$ years
  of the radio continuum and H$\alpha$ emission of HH~1 and 2. We find that the radio continuum
  and the H$\alpha$ emission of both objects show very similar trends, with HH~1 becoming fainter
  and HH~2 brightening quite considerably (about a factor of 2). We also find that the $F_{\rm H\alpha}/F_{ff}$ (H$\alpha$ to
  free-free continuum) ratio of HH~1 and 2 has higher values than the ones typically found in planetary nebulae (PNe),
  which we interpret as an indication that the H$\alpha$ and free-free emission of HH~1/2
  is produced in emitting regions with lower temperatures ($\sim 2000$~K) than the emission of PNe
  (with $\sim 10^4$~K).}

\keywords{shock waves --- stars: winds, outflows ---
Herbig-Haro objects --- ISM: jets and outflows --- ISM: kinematics and dynamics ---
ISM: individual objects (HH1/2) --- stars: formation}

\begin{document}
\maketitle

\section{Introduction}

HH~1 and 2 were the first discovered Herbig-Haro (HH) objects (Herbig 1951; Haro 1952).
Their diverging proper motions (Herbig \& Jones 1981) show that they correspond to the
two lobes of a bipolar outflow. This outflow was first thought to be ejected by the
Cohen-Schwartz (C-S) star, located closer to HH~1 (Cohen \& Schwartz 1979), but was
later shown to be ejected from the VLA~1 radio source (Pravdo et al. 1985), centrally
located between HH~1 and 2.

The radio continuum emission of HH~1 and 2 is clearly detected in maps obtained with the
Very Large Array (VLA) interferometer (Pravdo et al. 1985). The radio emission of these
objects shares the proper motions of their optical counterparts (Rodr\'\i guez et al. 1990).

The optical emission of HH~1 and 2 also shows relatively strong time variabilities (Herbig
1968, 1973). Raga et al. (2016a) used Hubble Space Telescope (HST) narrow-band images to show that
during the last $\sim 20$~years HH~1 has become fainter and HH~2 has brightened quite considerably.
In the present paper we show that the radio continuum emission of these objects shows
similar trends.

To this effect, we have generated a 4.86~GHz map using archival VLA observations
using a number of epochs centered around 1988 and a new map obtained with the 
Karl G. Jansky Very Large Array in 2012. These maps (as well as the HST H$\alpha$ images)
are described in section 2.

We carry out a comparison of the radio continuum and H$\alpha$ morphologies in section 3, and
calculate the angularly integrated emission in section 4. A comparison between the
radio continuum to H$\alpha$ ratios of HH~1 and 2 with the ones obtained for planetary
nebulae is made in section 5. Finally, the results are summarized in section 6.

\section{The observations}

\subsection{Very Large Array}

The first image of the HH 1/2 region was made with the Very Large Array (VLA) of NRAO\footnote{The National 
Radio Astronomy Observatory is a facility of the National Science Foundation operated
under cooperative agreement by Associated Universities, Inc.}  
at C-band (4.86 GHz) using data from 11 epochs between 1984 October 02 and 1992 December 19.
The parameters of these observations are
listed in Table 1 of Rodr\'\i guez et al. (2016).
The average epoch of these data is 1988.01.
These observations were all made with the phase center at or very close the position of
HH~1/2 VLA~1 [$\alpha(J2000) = 05^h~ 36^m~ 22\rlap.^s84$;
$\delta(J2000)$ = $-$06$^\circ~ 46'~ 06\rlap.{''}2$], the exciting source of the
HH~1/2 system (Pravdo et al. 1985; Rodr\'\i guez et al. 2000).
The data were calibrated following the standard procedures in the AIPS  (Astronomical Image Processing System)
software package of NRAO and then concatenated in a single file. 

The second image was made with the  Karl G. Jansky Very Large Array of NRAO in the C (4.4 to 6.4 GHz)
and X (7.9 to 9.9 GHz) bands
during 2012 May 26 (2012.40), under project 12A-240. The central frequency of the image is 7.15 GHz.
At that time the array was in its B configuration.  The phase center was 
at $\alpha(2000) = 05^h~ 36^m~ 22\rlap.^s00$;
$\delta(2000)$ = $-$06$^\circ~ 46'~ 07\rlap.{''}0$. The absolute amplitude calibrator was 0137+331 and
the phase calibrator was J0541$-$0541.
The digital correlator of the JVLA was configured at each band in 16 spectral windows of 128 MHz width each subdivided 
in 64 channels of 2 MHz. 
The total bandwidth of the observations was about 2.048 GHz in a full-polarization mode.
The data were analyzed in the standard manner using the CASA (Common Astronomy Software Applications) package of NRAO.

Both images were restored with the synthesized beam of the 2012.40 observations,
$1\rlap.{''}47 \times 0\rlap.{''}94; PA =  -24^\circ$,  and are shown in Figure 1.

\begin{figure}[!t]
\includegraphics[width=\columnwidth]{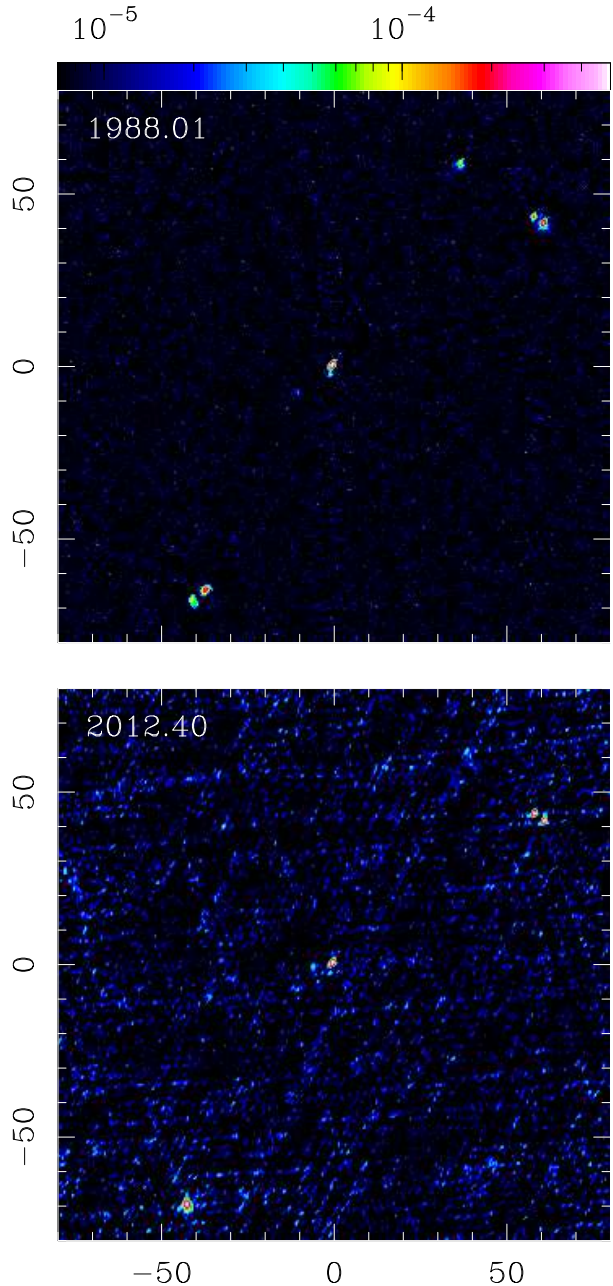}
\caption{VLA radio continuum maps at 4.86 GHz of the HH~1-2 region obtained in
  1988.01 (top) and at 7.15 GHz in 2012.40 (bottom). The axes are labeled with offsets (in arcsec)
  from the position of the VLA~1 outflow source (N is up and E to the left).
  HH~1 is the emission at the approximate position (40,60)$''$, and
  HH~2 at (-40,-65)''. The emission of the region around the Cohen-Schwartz
  star is seen in the 1988 map at (20,30)$''$. The maps are shown with the logarithmic
  colour scale given (in mJy per beam) by the top bar.}
\label{fig1}
\end{figure}

\subsection{Hubble Space Telescope}

We compare the VLA maps with the four epochs of HH 1/2 H$\alpha$ images
available in the HST archive:
\begin{itemize}
\item 1994.61: 3000s exposure (Hester et al. 1998),
\item 1997.58: 2000s exposure (Bally et al. 2002),
\item 2007.63: 2000s exposure (Hartigan et al. 2011),
\item 2014.63: 2686s exposure (Raga et al. 2015a).
\end{itemize}
The calibration of these images and the errors in the determined line fluxes
are described in detail by Raga et al. (2016a).

These images have been placed in approximately the same coordinate system
as the VLA maps by centering the positions of the emission of the near
environment of the Cohen-Schwartz star (visible in the H$\alpha$ images
and in the 1988.01 VLA map). This results in a $\sim 0\rlap.{''}2$ shift
of the H$\alpha$ images with respect to the positions derived from
an astrometric calibration obtained using the positions of the C-S star
and ``star number 4'' of Strom et al. (1985).

Figures 2 and 3 show the H$\alpha$ emission regions around HH~1 and 2 (respectively)
in the four available epochs. In these figures we show the shifting, circular
diaphragms (of $3''$ radius for HH~1 and $6''$ radius for HH~2) that we
have used to compute H$\alpha$ fluxes to compare with the free-free
radio continuum fluxes obtained from the VLA maps.

\begin{figure}[!t]
\includegraphics[width=0.97\columnwidth]{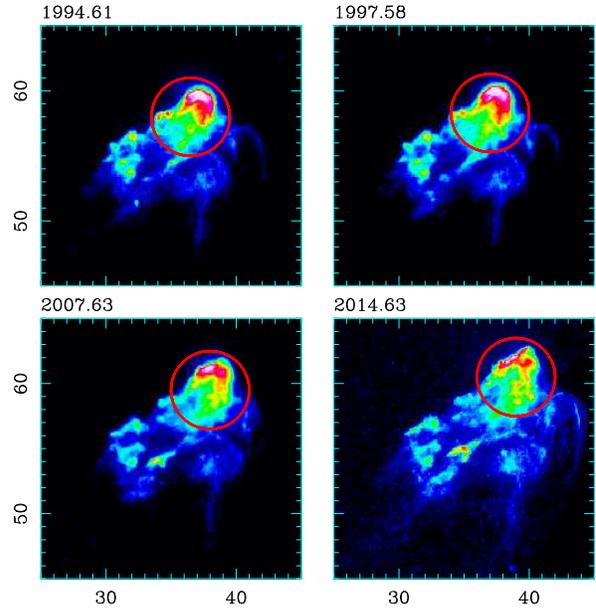}
\caption{H$\alpha$ images of HH~1 in the four available epochs of HST
  images. The axes are labeled as offsets (in arcsec) from the position
  of the VLA~1 outflow source. The circular diaphragms (of $3''$ radii) shown
  on the images have been used to compute H$\alpha$ fluxes. The images are displayed
  with a logarithmic colour scale.}
\label{fig2}
\end{figure}

\begin{figure}[!t]
\includegraphics[width=\columnwidth]{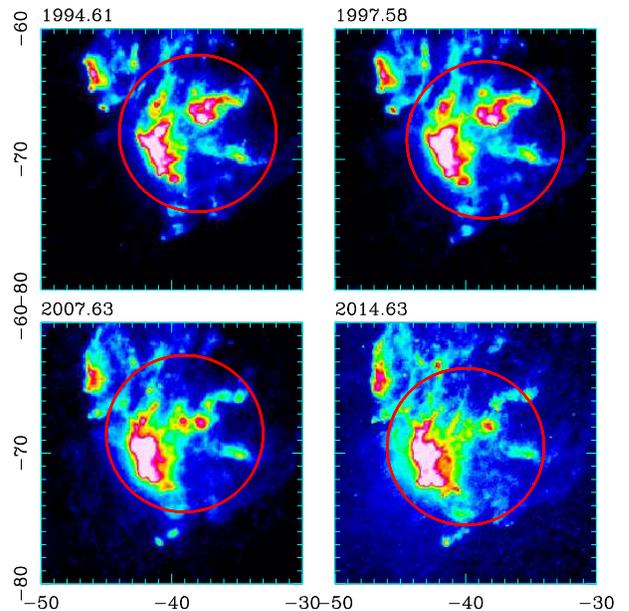}
\caption{H$\alpha$ images of HH~2 in the four available epochs of HST
  images. The circular diaphragms (of $6''$ radii) shown
  on the images have been used to compute H$\alpha$ fluxes. The images are displayed
  with a logarithmic colour scale.}
\label{fig3}
\end{figure}

\section{The free-free and H$\alpha$ emission}

Figures 4 and 5 show a comparison between the free-free continuum and the
H$\alpha$ emission of HH~1 and 2 (respectively). These figures show superpositions
of the 1988 VLA map and the 1997 H$\alpha$ image (top frames) and of the
2012 VLA map and the 2014 H$\alpha$ image (bottom frames of figures 4 and 5).

\begin{figure}[!t]
\includegraphics[width=\columnwidth]{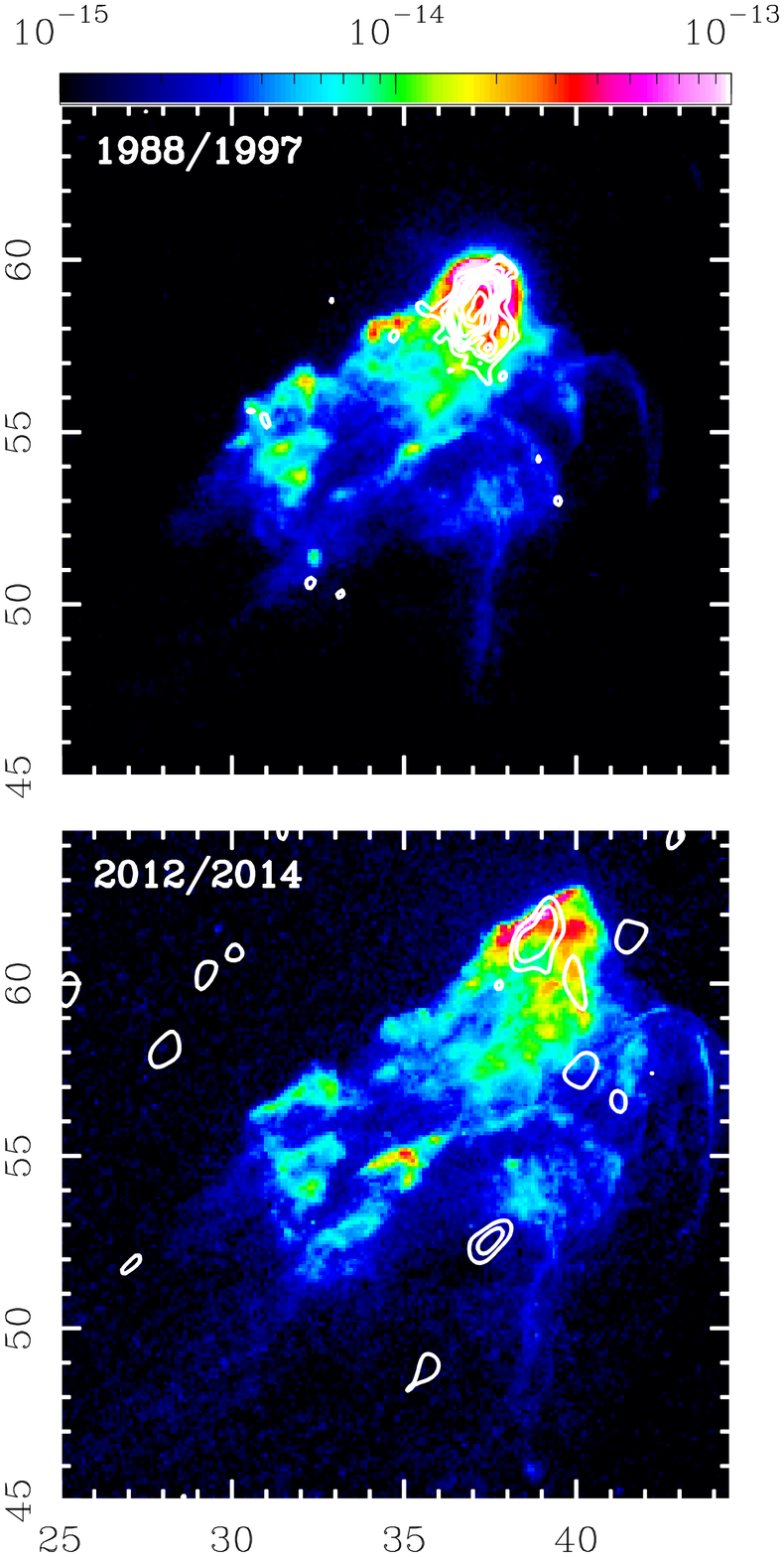}
\caption{Top: 1988 VLA map and 1997 H$\alpha$ image and bottom:
  2012 VLA map and 2014 H$\alpha$ image of HH~1. The VLA maps are displayed
  with linear contours (the bottom contour corresponding to 25
  and a linear step of 10 $\mu$Jy per beam). The H$\alpha$ maps are displayed with the
  logarithmic colour scale given (in erg cm$^{-2}$ s$^{-1}$) by the top bar.}
\label{fig4}
\end{figure}

\begin{figure}[!t]
\includegraphics[width=\columnwidth]{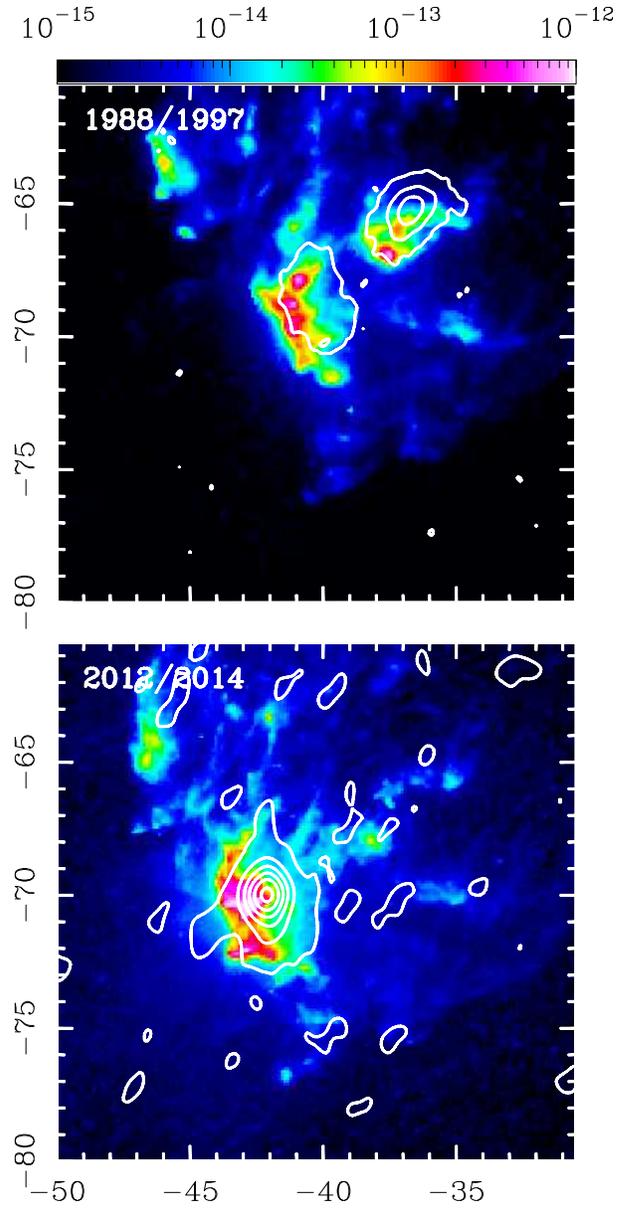}
\caption{Top: 1988 VLA map and 1997 H$\alpha$ image and bottom:
  2012 VLA map and 2014 H$\alpha$ image of HH~2. The VLA maps are displayed
  with linear contours (the bottom contour corresponding to 25
  and a linear step of 100 $\mu$Jy per beam). The H$\alpha$ maps are displayed with the
  logarithmic colour scale given (in erg cm$^{-2}$ s$^{-1}$) by the top bar.}
\label{fig5}
\end{figure}

For HH~1, we see that both the H$\alpha$ and free-free emission show a clear drop
between the first and second epochs (Figure 4). We also see shifts in the positions of the
radio continuum and H$\alpha$ emission peaks. These shifts are at least partly due
to the proper motions of HH~1 (of $\approx 300$~km~s$^{-1}$, see Raga et al. 2016b),
which correspond to $\sim 0\rlap.{''}3$ in the 2012-2014 time span (bottom frame)
and $\sim 1\rlap.{''}4$ in the 1988-1997 time difference (top frame of Figure 4) between
the VLA and the H$\alpha$ maps.

For HH~2, we see that the 1988 VLA map shows two separate condensations (H to the SE
and A to the NW, top frame of Figure 5). By 2012, condensation A has basically disappeared,
and condensation H has strengthened considerably (botton frame). A similar effect is seen in the
H$\alpha$ emission. In the comparison between the 2012 radio continuum and the 2014 H$\alpha$
emission we see a clear morphological difference, which probably cannot be fully attributed
to the proper motion of HH~2 (which corresponds to a shift of only $\sim 0\rlap.{''}3$ to the
SE between 2012 and 2014, see Raga et al. 2016c).

\section{The H$\alpha$ to free-free continuum ratios}

In Figure 6 we show the H$\alpha$ flux within the diaphragms shown in Figures 2 and 3
(for HH~1 and 2, respectively). This figure also shows the radio continuum flux
(integrated over diaphragms of the same sizes as the ones used for H$\alpha$) in the
two available epochs.

\begin{figure}[!t]
\includegraphics[width=\columnwidth]{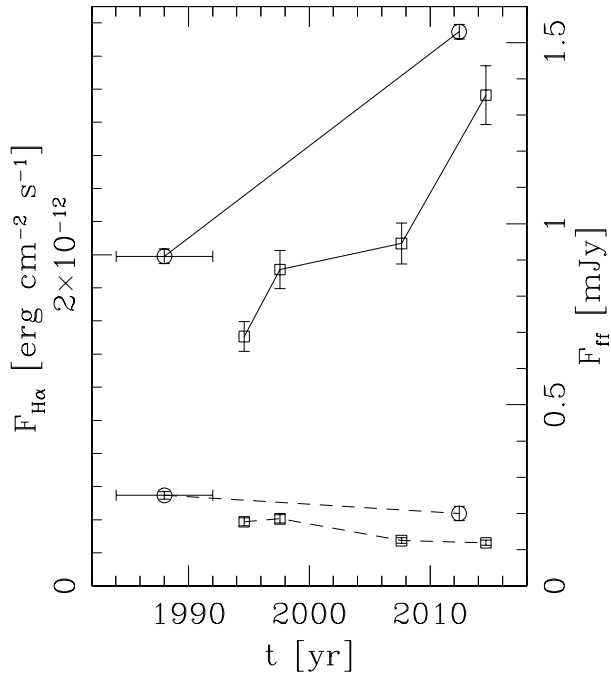}
\caption{H$\alpha$ fluxes (open squares) and radio continuum fluxes (open circles)
  as a function of time. The horizontal bars indicate the time interval over which the
1988 image was obtained. The HH~1 fluxes are joined by dashed lines, and the HH~2 fluxes
  are joined by solid lines. The scale of the H$\alpha$ fluxes is given on the left axis
  and the scale of the radio continuum fluxes is given on the right axis.}
\label{fig6}
\end{figure}

It is clear that the H$\alpha$ and radio continuum fluxes both show an increasing
flux vs. time trend for HH~2, and a decreasing trend for HH~1. Within the errors,
the observed trends (in the radio continuum and in H$\alpha$) are similar for
both HH~1 and 2.

In order to estimate the ratio between the H$\alpha$ flux and the free-free continuum,
we use the 2014 H$\alpha$ and the 2012 continuum fluxes, because
they are the pair of values closer in time. These two fluxes and their ratios
are given in Table~1 (for HH~1 and 2). Using the 1997 H$\alpha$ and 1988 radio
continuum fluxes (see Figure 6), one obtains similar line to continuum ratios.

\begin{table}[!t]
\small
\caption{H$\alpha$ and free-free fluxes and ratios}
\begin{tabular}{lcc}
\hline
\hline
& HH~1 & HH~2 \\
\hline
$F_{ff}$$^1$ &  $0.20 \pm 0.02$ & $1.53 \pm 0.02$ \\
$F_{\rm H\alpha}$$^2$  & $2.60 \pm 0.16$ & $29.6\pm 1.8$ \\
$F_{\rm H\alpha}/F_{ff}$$^3$ &  $1.30\pm 0.09$ & $1.94\pm 0.14$ \\
$F_{\rm H\alpha,0}/F_{ff}$$^{3,4}$ & $2.34\pm 0.16$ & $3.50\pm 0.25$ \\
\hline
\hline
\multicolumn{3}{l}{$^1$ free-free fluxes in mJy}\\
\multicolumn{3}{l}{$^2$ observed H$\alpha$ fluxes in $10^{-13}$~erg~s$^{-1}$~cm$^{-2}$}\\
\multicolumn{3}{l}{$^3$ ratios in $10^{-12}$~erg~s$^{-1}$~cm$^{-2}$mJy$^{-1}$}\\
\multicolumn{3}{l}{$^4$ dereddened free-free/H$\alpha$ ratio}\\
\end{tabular}
\end{table}

These line-to-continuum ratios are most interesting. In order to compare them
with theoretical predictions of this ratio, we should correct the observed
values for interstellar extinction. As discussed by Raga et al.
(2016), for an $E(B-V)=0.27$ and a standard Galactic extinction curve in order
to obtain the dereddened H$\alpha$ flux, one has to multiply the observed flux
by a factor of 1.80. The resulting, dereddened line to continuum ratios
(calculated with the 2012 VLA map and the 2014 H$\alpha$ image) are
given in Table~1.

We can compare the dereddened $F_{\rm H\alpha,0}/F_{ff}$ values obtained for HH~1 and 2
with the prediction for this ratio obtained from equation (2) of Reynolds (1992).
The predicted temperature dependence for this ratio is shown in Figure 7. In this
Figure, we also show horizontal lines corresponding to the dereddened ratios obtained for
HH~1 (short dashes) and HH~2 (long dashes). It is clear that the observed ratios
would imply that the emission has a dominant contribution from regions with
$T\sim 1000\to 3000$~K.

\begin{figure}[!t]
\includegraphics[width=\columnwidth]{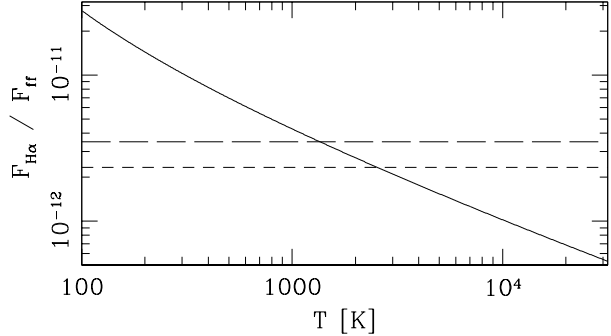}
\caption{The solid line shows the predicted ratio of the recombination H$\alpha$ flux
  (in erg cm$^{-2}$ $s^{-1}$) to the 4.86 GHz free-free continuum (in mJy) as a function
  of temperature. The large dash and small dash lines show the values of the dereddened
  ratios found for HH~1 and 2 (respectively).}
\label{fig7}
\end{figure}

\section{Comparison with the radio continuum and H$\alpha$ emission of PNe}

To determine observationally the expected $F_{\rm H\alpha}/F_{ff}$ ratio for photoionized
nebulae, we used the catalogs of Frew et al. (2013) and Parker et al. (2016)
to select planetary nebulae with the following criteria:

i) determined 6-cm flux density with value $\geq$1 mJy,

ii) determined H$\alpha$ flux, 

iii) determined logarithmic extinction at H$\beta$, $c_\beta$, from which the 
logarithmic extinction at H$\alpha$ can be obtained (Frew et al. 2013) as

$$c_\alpha = 0.70 \times c_\beta.$$

Planetary nebulae are the better objects for this determination since
HII regions can suffer from very large extinction.
We found a total of 211 planetary nebulae that comply with the above criteria. 
In Figure 8 we plot their extinction-corrected H$\alpha$ flux as a function of their
6-cm flux density. We fitted these data points with a linear function with slope 1. The least-squares fit gives

$$\log_{10}F_{\rm H\alpha} = -(12.0 \pm 0.1)~ +~\log_{10} F_{\rm 6cm}\,,$$

\noindent where $F_{\rm H\alpha}$ is given in erg cm$^{-2}$ s$^{-1}$ and $F_{6cm}$ is given in mJy.
This fit suggests (see Figure 7) that the $H_\alpha/H_{ff}$ ratio in planetary
nebula can be explained on the average as coming from photoionized gas at a temperature of
$\sim 10^4$ K.

In the same Figure we show the fluxes of HH1 and HH2, and it can be seen that
their $F_{\rm H\alpha}/F_{ff}$ ratios are a factor of $\sim$2 to 4 larger that the average value for
planetary nebulae. These departures from the mean are, however, not very significant given the high dispersion of the
planetary nebulae data. The mean and standard deviation of $log_{10}(F_{\rm H\alpha}/F_{\rm 6cm})$ for
the planetary nebulae
are -12.0$\pm$0.3 and thus the HH objects are separated from the planetary nebulae mean only by
1-2 standard deviations (see Table 1).

We note that we have not taken into account two small effects. On one hand, the free-free
emission can have a contribution from ionized helium. This could introduce an extra contribution
of the order of 10\%\ to the
free-free emission from pure hydrogen. On the other hand, the planetary nebulae radio
data was taken at 6-cm (5 GHz), while the points shown for HH1 and HH2 were taken at 7.15 GHz.
Assuming that we are observing optically thin free-free, the flux density is
expected to go as $\nu^{-0.1}$ and this will introduce an underestimate in the 6-cm flux density of
the HH objects of about 4\%.

\begin{figure}[!t]
\includegraphics[width=\columnwidth]{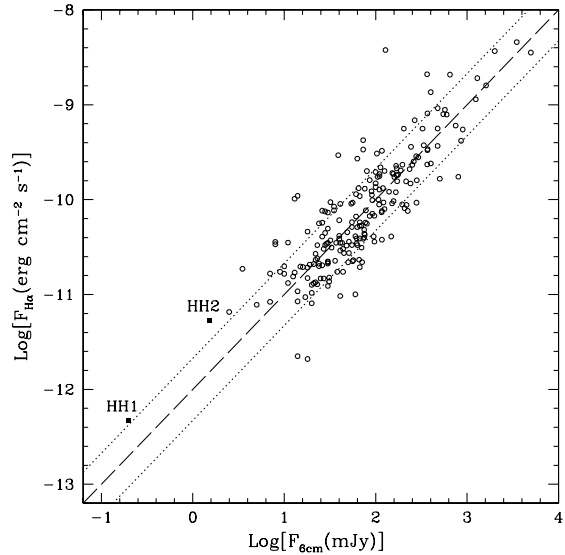}
\caption{H$\alpha$ flux versus 6-cm radio continuum flux for 211 planetary nebulae (empty circles).
The dashed line shows the least squares fit to these points as discussed in the text. The dotted lines show the $\pm$1 $\sigma$ ($\sigma$ = 0.33)
confidence interval. The positions of HH1 and HH2 are labeled and shown as filled squares.}
\label{fig8}
\end{figure}

The difference between the $F_{\rm H\alpha}/F_{ff}$ ratios of PNe and of HH~1/2 is significant and a comparison with a larger sample of
HH objects could be interesting. The straightforward explanation of this difference is that while all
photoionized regions (in particular, PNe) have temperatures $\sim 10^4$~K (resulting from
the balance of the photoionization heating and the strongly rising forbidden line cooling,
see e.g. the book of Osterbrock 1974), the cooling region behind shock waves has emission
at a range of decreasing temperatures. Particularly the H$\alpha$ emission (as well
as the free-free emission) has a strong contribution from the dense, $T\sim 10^3$~K
region towards the trailing edge of the recombination zone (this effect is discussed
by Raga \& Binette 1991, but is present in all plane-parallel shock models). Therefore,
the fact that the $F_{\rm H\alpha}/F_{ff}$ values of HH~1/2 imply a gas temperature of
$\sim 2000$~K (see Figure 7) is not surprising.

Another effect that could be affecting the HH~1/2 $F_{\rm H\alpha}/F_{ff}$ ratio
is that collisional excitation of H$\alpha$
appears to be taking place in part of the emitting regions of these objects (Raga et al. 2015b, c).
However these regions have small angular extents, and do not contribute substantially
to the angularly integrated emission of HH~1 and 2 (Raga et al. 2016a).

\section{Summary}

We have presented a comparison of two VLA-JVLA radio continuum maps (epochs
1988 and 2012) with four HST H$\alpha$ images (1994, 1997, 2007 and 2014) of HH~1 and 2.
We find that in both the radio continuum and H$\alpha$ images:
\begin{itemize}
\item HH~1 shows a trend of decreasing intensities with time,
\item HH~2 shows a general trend of increasing intensities, with condensation H
  becoming much brighter and condensation A fading away.
\end{itemize}
The fact that both the radio and the optical emission show similar trends with time
is quite conclusive evidence that the time-evolution of HH~1 and 2 is not due to
a variation of the extinction (which could occur if the HH objects are moving into
or away from regions with higher extinction). A change with time of the extinction
towards the moving objects would affect the optical, but not the radio emission.
This result agrees with Raga et al. (2016a) who reached a similar
conclusion from an analysis of the time-dependence of the optical/UV
emission line spectra of HH~1 and 2.

We find that the ratio $F_{\rm H\alpha}/F_{ff}$ between the (angularly integrated) H$\alpha$ and free-free
continuum fluxes of HH~1/2 agrees with the theoretical prediction obtained for a $T\sim 2000$~K
emitting gas. This ratio is considerably higher than the one predicted for a $10^4$~K
temperature.

This effect shows up as a significant difference between the $F_{\rm H\alpha}/F_{ff}$ values of
HH~1/2 and the typical values obtained for a selection of PNe (with measured radio and H$\alpha$ fluxes),
which on the average do have $\sim 10^4$~K temperatures, as expected for photoionized regions.
This leads us to suggest that the value of $F_{\rm H\alpha}/F_{ff}$ is an interesting diagnostic
that can be used to discriminate between HH objects and photoionized regions. However, there appear to be a significant fraction of planetary nebulae
as cool as the HH objects and this issue deserves further research.

\acknowledgments
We are thankful to R. Estalella for his valuable
comments. This research has made use of the HASH PN database at hashpn.space.
ARa acknowledges support from the CONACyT grants
167611 and 167625 and the DGAPA-UNAM grants IA103315, IA103115, IG100516 and IN109715.
LFR acknowledgs the support from CONACyT, M\'exico and DGAPA, UNAM.

\end{document}